\documentstyle[pre,aps]{revtex}
\begin{document}

\def\half{{1\over 2}}
\def\tr{{\rm tr}}
\def\e{{\rm e}}
\def\i{{\rm i}}
\def\eps{\epsilon}
\def\qhat{{\hat Q}}
\def\xhat{{\hat x}}
\def\yhat{{\hat y}}
\def\qtil{{\tilde Q}}
\def\stil{{\tilde S}}
\newcommand{\bra}[1]{\langle#1 |}
\newcommand{\ket}[1]{| #1\rangle}
\draft
\title{Response Function of an Irregular Oscillator} 

\author{Hirokazu Aiba$^1$~,~ and ~Toru Suzuki$^2$}

\address{$^1$Koka Women's College, 38 Kadono-cho Nishikyogoku, Ukyo-ku,
         615-0882 Kyoto, Japan\protect\\
     $^2$Department of Physics, Tokyo Metropolitan University, 
          192-0397 Hachioji, Japan}
\vspace{1cm}

\date{\today}

\maketitle

\begin{abstract}
      Properties of the response functions for a 
      two-dimensional quartic oscillator 
      are studied based on the diagonalization of the 
      Hamiltonian  in a large model space. In particular, 
      response functions corresponding to a given 
      momentum transfer are studied for different values 
      of the coupling parameter in the Hamiltonian. The latter 
      controls regular or chaotic nature of the spectra and 
      eigenstates of the system.  Fluctuation properties of the 
      energy-strength correlation of the response  are  
      investigated. Even when the statistical properties of the 
      system indicate an almost completely chaotic character, 
      there remains a typical structure in the response function 
      similar to that in the regular system. The nature of this 
      structure is studied in some detail.
\end{abstract}

\section{Introduction}
\label{sec:intro}

  Quantum mechanical  manifestations of  dynamical properties of a 
system which classically possesses a chaotic character have been 
intensively studied\cite{berry,bohigas}.
Level statistics which has a long history in nuclear physics 
as described by the random-matrix theory\cite{brody,weiden} is 
now  a favorable playground 
in the discussion of a transition from a regular (integrable) to a 
chaotic character of a quantum system.  Together with numerical 
studies on  model systems, analytic investigation has been made 
based on semiclassical trace formula\cite{gutz}. Wave functions of 
a system which is classically chaotic have also been investigated: 
Statistical theory predicts that the amplitude distributions show 
the Porter-Thomas distribution\cite{porter}, which was then numerically 
demonstrated to hold  for model chaotic systems.  Contrary to the 
naive expectation, however, the profile of the wave function 
for a chaotic system is not entirely structureless: For instance, 
the Husimi representation of a wave function in the chaotic regime 
frequently suffers from a {\it scar} of classical 
periodic orbits\cite{heller}. Although a considerable progress 
has since been made, it is still an important issue to 
clarify the characteristic of eigenstates and its matrix elements 
for systems which are classically irregular or chaotic. 

  It is the purpose of the present paper to study another aspect of 
the wave functions for a system which shows a transition from an 
integrable to a chaotic character: We study response 
functions of the system, i.e. 
transition matrix elements as a function of energy. 
Statistical properties of the distribution of transition 
matrix elements have been studied\cite{Feingold,Alhassid,Wilkinson}, 
and it was shown, in particular, that the distribution becomes a 
Porter-Thomas type for chaotic systems. The  approach proposed in 
Ref.\cite{Wilkinson} has since been developed to elucidate the role 
of periodic orbits and was extended to various systems 
including the response of mesoscopic systems to realistic 
probes\cite{eckardt,MW,mehlig,kon}. These studies are based on 
the semiclassical framework, and focus mainly on the  responses to 
long-wavelength probes. Semiclassical studies of response functions 
have also been done in a different framework\cite{schuck}, which 
concentrate on their smooth behavior but not much on fluctuations.

In this paper, we study a model system using a large space 
diagonalization and calculating response functions for the 
operators which probe the 
system with a variable wavelength, or momentum transfer.
In particular, we put an emphasis  on which aspect of the response
function reflects the regular or the chaotic character of the system.
We also would like to study a structure  in the response function 
for a system in a chaotic regime which however is not expected to 
occur in the statistical random-matrix theory. 
Although we are here concerned with the properties of 
response functions which show up in a model system, they  will also be 
interesting in  realistic applications, as this response is similar 
to the excitation cross section for, e.g., electron scattering 
 in the plane-wave Born approximation. 
Thus it is hoped that the present study may  provide insight 
into the understanding of collective states such as nuclear 
giant resonances embedded in complicated many-particle 
many-hole states as studied in nuclear reactions. 

The main content of the paper is as follows: In the next section 
we summarize classical and quantum mechanical properties of 
the model Hamiltonian and fix values of the relevant parameters. 
In Sec.\ \ref{sec:response} we study response functions 
first for a long wavelength probe, and then for the probe  
characterized by a given momentum transfer. We  study the fluctuation 
properties of the response functions, concentrating especially on 
the similarity or the difference for the regular and the chaotic 
 systems.  The accuracy of the calculation has been checked against 
 sum rules. Final section is devoted to a summary.

\section{Basic Ingredients of the Model}
\label{sec:model}
  
   In order to study the response functions for a 
system which is capable of showing regular as well as chaotic properties, 
we adopt the following Hamiltonian as a model,
\begin{equation}
   H=\half(p_x^2+p_y^2)+\half(x^4+y^4)-kx^2y^2.  
\label{hamiltonian}
\end{equation}
This model Hamiltonian has been adopted by a number of 
authors for the studies of level statistics or wave 
functions\cite{Percival,Meyer,Zimmer,Pollak,Boh}. 
It was also employed as a model for a background system in the studies 
of the fluctuation properties of strength functions.\cite{ams} 
Let us first briefly summarize 
classical properties of the model, which have been 
studied in detail by Meyer\cite{Meyer}.  The Hamiltonian (\ref{hamiltonian}) 
possesses a dynamical scaling property in the sense that the 
classical phase space structure at one energy is mapped into 
another by a simple scaling of the coordinates and momenta.  
It has a high symmetry called $C_{4v}$: 
The Hamiltonian is invariant with respect to  
a reflection about $x$-axis, $y$-axis, and also about the 
line $x=y$.  Furthermore, by rotating 45$^{\circ}$ in the 
$x-y$ plane, the Hamiltonian is mapped into the one with a 
coupling constant $k'=-(3+k)/(1-k)$. As the system becomes 
unbounded from below for $k>1$, we have only to consider 
the range $[-1,1]$ for the coupling constant $k$.  Meyer\cite{Meyer} 
showed that for large $k$ values ($\ge 0.4$) the classical 
phase space structure is almost completely chaotic, while for 
small $k$ the system becomes regular.  In the following calculations 
 we adopt two typical values of the parameter: $k=$0.2 and 0.6. They 
correspond to quasiintegrable and fully 
chaotic systems, respectively. For instance, for $k=0.6$ a 
single trajectory fills up almost 90\% of the 
available phase space, while for $k=0.2$ the fraction of the phase 
space covered by irregular orbits is only 25\% in typical 
cases\cite{Meyer}. 

  To study quantum mechanical properties of the eigenstates 
of the Hamiltonian (\ref{hamiltonian}), we follow the procedure essentially 
 of Zimmerman et al.\cite{Zimmer}:  The Hamiltonian is 
diagonalized within a truncated model space spanned by a set of 
suitable harmonic oscillator bases $\ket{n_x,n_y}$, 
where $n_x, n_y$ denote the numbers 
of oscillator quanta in the $x$- and $y$-directions. In 
the following we take the unit $\hbar=1$. The frequency 
$\omega_0$ of the harmonic oscillator basis 
is determined so as to minimize $\tr H$ in the adopted model space. 
The obtained values of $\omega_0$ are 7.51($k=0.2$) 
and 7.13($k=0.6$). 
The Hamiltonian matrix can be decomposed into submatrices due to 
the $C_{4v}$ symmetry. As in Ref.\cite{Pollak} we take up four 
classes of the one-dimensional representation which are labeled as 
$A_1, A_2, B_1, B_2$ according to their symmetry properties 
under reflection on the axes and diagonals in the $x-y$ 
plane \cite{Meyer}. (For instance, $A_1$ is symmetric under both 
reflections.)  The model space is spanned by the bases with  
$0\leq  n_x+n_y\leq  300$,  which gives the 
dimension of each submatrices as 5776, 5625, 5700, 5700. The 
diagonalization has been performed for each submatrices.  
Study of the nearest-neighbor spacing distribution confirms the 
character of the system suggested by the classical phase space
structures, i.e., the Poisson like distribution for $k=0.2$ and 
the Wigner distribution for $k=0.6$ within each symmetry class. 
We also confirmed that the amplitude distribution of 
the wave functions for $k$=0.6 show the Porter-Thomas distribution 
except for a singular peak at zero.  

In the following the results of the calculation will be shown for the 
states which belong to the symmetry class $A_1$. The results are similar 
for other symmetry classes. 

The basis state $|n_1,n_2)_{\rm ES}$ belonging to the class $A_1$
is written as
\begin{equation}
|n_1,n_2)_{\rm ES}=\sqrt{\frac{1+\delta_{n_1n_2}}{2}}
(\ket{n_1,n_2}+\ket{n_2,n_1}),
\label{basis}
\end{equation}
where $n_1$ and $n_2$ are even integers and $n_1\le n_2$.
Because of the selection 
rule, the relevant part of the operator which contributes to the matrix 
element of the response should have the definite symmetry property. 
It should be noted  that among the 5776 wave functions of the class 
$A_1$, those having very large energies are not quite reliable because of the 
limitation in the basis states. This is especially so when  
the response functions 
with large momentum transfer $q$ are concerned. We may make an estimate 
for the range of validity by comparing the obtained level density 
with the semiclassical one. The comparison suggests that the maximum reliable 
energy to be $E=1000\sim 1500$ depending on the values of the parameter $k$ 
in Eq.\ (\ref{hamiltonian}). This maximum energy is contrasted 
with the  largest energy eigenvalue 
$E\simeq 3000$ obtained by the diagonalization. This limits the maximum 
value of the momentum transfer of the probe adopted below 
to be around $q\simeq 50$, where the 
corresponding \lq quasielastic peak' lies around 
$E_{\rm peak}=q^2/2\simeq 1250$. This 
is confirmed by the calculation as shown later. 

\section{Response Functions}
\label{sec:response}

We consider the response functions defined by 
\begin{equation}
  W^{(i)}(\Omega)\equiv\sum_j|\bra{j}\qhat\ket{i}|^2
                 \delta (\Omega-(E_j-E_i)) ,  
\label{response}
\end{equation}
where $\qhat$ denotes a probing operator which connects 
the initial and the final eigenstates $\ket{i}$ and $\ket{j}$. 
In many cases of interest, the initial state $\ket{i}$ is 
set to the ground state of the system $\ket{gs}$
 which belongs to the $A_1$ symmetry class, in which 
case the index $(i)$ is dropped. The response function shows the 
distribution of the state $\qhat\ket{i}$ over 
the energy eigenstates $\{\ket{j}\}$. If, for instance, the initial state 
$\ket{i}$ has a simple structure as in the ground state of the harmonic 
oscillator, the state $\qhat\ket{i}$ and the response function will 
simply reflect the structure of the probe operator $\qhat$. 
We consider operators depending 
only on a single variable, say $\xhat$, to see how the irregular behavior 
of the wave functions controlled by the parameter $k$  may be 
reflected  in the response function. 
One may rewrite the response function (\ref{response}) in the form of the 
time-correlation function 
\begin{equation}
  W^{(i)}(\Omega)={1\over 2\pi}\int_{-\infty}^\infty\!\! dt\, 
  e^{i\Omega t}\,
  \langle \qhat(\xhat(t))^\dagger\qhat(\xhat(0))\rangle_i, 
\label{timecorr}
\end{equation}
where $\langle$ $\rangle_i$ denotes the expectation value in the 
initial state $\ket{i}$. The probe $\qhat$ is written as a function 
of the operator 
\begin{equation}
    \xhat(t)\equiv e^{iHt}\xhat\,e^{-iHt} = \xhat + {\hat p}_x t
      -(\xhat^3-k\xhat{\hat y}^2)t^2+\cdots,
\label{operator}
\end{equation}
which is the solution of the Heisenberg equation of motion. 
In Eq.(\ref{operator}) we show also a short time expansion in terms of 
the operators at $t=0$, e.g., 
$\xhat(0)=\xhat$. A corresponding semiclassical 
expression for Eq.\ (\ref{timecorr}) has been fully utilized in the analysis 
of Refs.\cite{Wilkinson,eckardt,MW,mehlig,kon}. 

 It is generally believed that the universal behavior of a dynamical 
system, i.e., if it is regular or chaotic, emerges in the fluctuation 
properties of the  matrix elements of the operators, while their 
expectation values are strongly dependent on the specific 
dynamics of the system. Although the transition matrix 
elements for a chaotic system are known to generically follow the 
Porter-Thomas distribution, the energy-strength correlation 
such as the one contained in response functions is certainly 
dependent on the specific properties of the dynamics governed 
by the Hamiltonian. In this latter 
respect we note that the shape of the response function 
versus energy is constrained by a number of sum 
rules\cite{sum}.   Let us define 
\begin{equation}
   S^{(i)}_n (\qhat)= \int_{-\infty}^\infty \!d\Omega 
     ~\Omega^n W^{(i)}(\Omega)=\sum_j(E_j-E_i)^n
     |\bra{j}\qhat\ket{i}|^2  
\label{sumrule}
\end{equation}
for a given operator $\qhat$. The integer $n$ may in general 
take negative values (usually for $i=g.s.$ and $j\ne g.s.$), in which case 
the sum rule corresponds to the 
generalized susceptibility of the system. By increasing the value of 
$n$ and subtracting the 
lower moments, e.g., in the form of the shifted moment or of the 
cumulant, one can regain finer and finer structure of the response function. 
Although one may  recover the response functions 
by Mellin-transforming the sum rule values, the use of  sum 
rules lies in the fact 
that in some cases the sum becomes a simple matrix element in the initial 
state. The latter may be calculated precisely and serves as a check for the 
accuracy of the calculation. The low $n$ sum rules, 
in particular, sometimes become insensitive to the detailed dynamics  
and constrain the gross behavior of the response functions. 

 As a probe $\qhat$ of the response we first  consider 
the operator $\xhat^2$ with an arbitrary initial state $\ket{i}$. We then 
fix  $\ket{i}=\ket{gs}$ 
and adopt the operator $\qhat_q\equiv\e^{\i q\xhat}$ which is closely 
related to an excitation of the system by an external probe characterized by 
momentum transfer $q$ (and length scale $1/q$). 
The operator  $\xhat^2$ may be regarded as 
a long wavelength part of $\qhat_q$, and is similar 
to the $E2$ operator of electromagnetic transitions. 
By changing the value of $q$ in $\qhat_q$, one can study in principle 
long as well as short distance structure of the matrix elements.  

In the actual calculation we consider only the operators symmetric under 
the reflection about $x$- and $y$-axes and also about the line $x=y$, i.e., 
\begin{equation}
    \qtil \equiv (\qhat)_{A_1}={1\over 4}
           (\qhat(\xhat)+\qhat(-\xhat)+
            \qhat({\hat y})+\qhat(-{\hat y})), 
\label{qtild}
\end{equation}
where the arguments are explicitly written to show the dependence on 
the coordinates. 
For the initial state $\ket{i}$ in the class $A_1$ this 
implies that in Eq.\ (\ref{response}) only 
the states $j$ belonging to the $A_1$ symmetry class  contribute. 

\subsection{Response to the $\xhat^2$ Probe}
\par
We first  study the response function for the $\xhat^2$ probe. A typical 
example of the response to this probe is given in Fig.\ref{Respx}. Here 
we show the response $W^{(i)}(\Omega=E-E_i)$ for 
the $i$=500th initial state as a function of the 
energy $E$ for $k=$0.2 and 0.6. 
For other initial states the main features are similar. 
We immediately see that the response consists of three clusters of 
strengths for both $k$=0.2 and 0.6: Largest strength lies at $E=E_i$ with 
almost no other strength close to this peak, while two other clusters 
are located around $E=E_i\pm\Delta$, where 
$\Delta$ is slightly less than $2\omega_0$, the expected value 
for a simple harmonic oscillator. For  $k=0.2$, the strengths
are concentrated on a few states, while strengths are distributed
over some energy range for  $k=0.6$. We now introduce 
creation and annihilation operators $a_x^\dagger, \, a_x$, etc., of 
the oscillator quanta with frequency $\omega_0$, and decompose 
the operator ${\tilde {x^2}}$  as 
\begin{equation}
{\tilde {x^2}}=D_0+D_2^\dagger+D_2; \quad
     D_0\equiv\frac{1}{2\omega_0}({\hat n}_x+{\hat n}_y+1),\quad
     D_2^\dagger\equiv\frac{1}{4\omega_0}({a_x^\dagger}^2+{a_y^\dagger}^2).
\label{x2}
\end{equation}
One may then be tempted to assign the $E=E_i$ peak to the response to 
the operator $D_0$, 
and two other clusters around $E=E_i\pm\Delta$ to the operators 
$D_2^\dagger$ and $D_2$, respectively.
Figure\ \ref{resd0d2} (a) and (b) show the response  to
the operator $D_0$ for $k=0.2$ and 0.6, respectively.
Although the state $D_0\ket{i}$ is not proportional to $\ket{i}$, the 
fragmentation of the strength is restricted to only a few states in both 
cases.  More important is the mixing in the state $D_2^\dagger\ket{i}$.  
Figure\ \ref{resd0d2} (c) and (d)
show the response to the operator $D_2^\dagger$ for the $i=500$th 
state. The distribution of strengths
is considerably different between $k=0.2$  and $k=0.6$ cases.
Strengths for $k=0.2$ are seen to concentrate on a few states,
while those for $k=0.6$ are distributed over many states.

The above features may be quantified by studying the number of 
principal components (NPC). The NPC for a normalized state $\ket{\alpha}$ 
in terms of a complete set of orthonormalized states $\{\ket{j}\}$ is 
defined as:
\begin{equation}
N_{\rm pc}^{(\alpha)}\equiv(\sum_j (\langle j|\alpha \rangle)^4)^{-1}.
\label{npc}
\end{equation}
The NPC becomes unity when the strengths are concentrated in a single
eigenstate, while becomes $N_{\rm tot}$ when the strengths are equally
distributed over the whole $N_{\rm tot}$  eigenstates. 
Figure\ \ref{d0npc} shows the NPC of the state $D_0\ket{i}$ (with a 
suitable normalization) for each eigenstate $\ket{i}$ as a function 
of $E_i$, where the set $\{\ket{j}\}$ has been taken to be the 
eigenstate of the Hamiltonian. The NPC takes values $1\sim 1.5$ in 
most cases and 
decreases as the energy increases. General trend is not much different for 
$k=0.2$ and $k=0.6$. 
The situation is drastically different if we study the NPC for the 
state $D_2^\dagger\ket{i}$ (again normalized) as seen in Fig.\ \ref{d2npc}.
For $k=0.2$, the NPC remains small and does not show a marked energy 
dependence, while for  $k=0.6$,
the NPC shows a rapid increase as a function of energy and takes a 
quite large value. We thus find that the $D_2(D_0)$ part of the operator 
${\tilde {x^2}}$ is sensitive(insensitive) to the characteristic changes 
in the dynamics governed by the parameter $k$. 

There is another measure to see the difference between the two cases, 
$k=0.2$ and $k=0.6$, which can be obtained from the response function
associated with ${\tilde {x^2}}$. In Fig.\ \ref{Strength} the fraction 
of strengths (omitting the one for $E=E_i$) 
exhausted by  two major states carrying largest strengths 
for each initial state $\ket{i}$ 
is plotted against $E_i$. For $k$=0.2 more than 60\% of the total 
strengths is exhausted by the two major states and the distribution of
the fraction is almost 
independent of the initial energy $E_i$,  
while for $k=0.6$ they carry less than 50\% and this fraction 
decreases as a function of the initial state energy. Since the 
dominant part of the strength associated with the operator $D_0$ is 
contained in the initial 
state $\ket{i}$ and is omitted here, Fig.\ \ref{Strength} 
shows the characteristics for the response  to the operator 
$D_2^\dagger$ (and $D_2$) in accordance with the results from NPC. 

These studies imply that it depends strongly on the choice of the probe
 whether the difference in the character of the dynamics, namely regular 
 or chaotic, may be easily seen in the response function.
In the present case, the difference in the dynamics is not apparent for 
the probe $D_0\sim x^2+p_x^2$, while it becomes quite significant for 
$D_2\sim x^2-p_x^2$ which is a probe, in a sense, \lq orthogonal\rq \, to 
the unperturbed oscillator Hamiltonian $\sim D_0$. 
The response function for the probe 
${\tilde x^2}$ shows both  characteristics. 

\subsection{Response to the Probe $\qhat_q$}

We now consider the response function for $\qhat_q=\e^{\i q\xhat}$. 
We fix here the initial state to be the ground state. In this case the 
response is closely related to the  situation of physical interest such as 
the inelastic electron scattering from the target in the ground state, where 
$q$ gives the momentum transfer on the target. The 
symmetrized probe for $\qhat_q$ is given by 
            $\qtil_q\equiv (\qhat_q)_{A_1}= 
            {1\over 4}(\e^{\i q\hat x}+\e^{-\i q\hat x}+\e^{\i q\hat y}
            +\e^{-\i q\hat y})$. 
The response functions $W(q,\Omega\equiv E-E_{gs})$ (for $\qhat_q$) and 
${\tilde W}(q,\Omega)$ (for $\qtil_q$) are calculated in terms of the 
elementary matrix element $Q_{nm}(q)\equiv \bra{n} 
\e^{\i q\hat x}\ket{m}$ for a
one-dimensional harmonic oscillator between the states with quanta $n$ 
and $m$, which is given by 
\begin{equation}
             Q_{nm}(q)=i^\alpha\e^{{1\over 2}z}
                            z^{{1\over 2}\alpha}
                        \sqrt{n!\over (n+\alpha)!}L^\alpha_n(z)
                        \hskip 1cm (m\geq n) , 
\label{qelement}
\end{equation}
where $\alpha\equiv m-n$, $z\equiv q^2/2\omega_0$ and 
 $L^\alpha_n(z)$ denotes the associated Laguerre polynomial. The functions 
 $Q_{nm}(q)$ are  calculated from recursion relations.

We consider several $q$ values corresponding to different 
resolution of the probe, the small $q$ limit being related 
to the long-wavelength probe $\xhat^2$ above. On the other 
hand, for large $q$ values the operator resolves a fine structure 
of the system and the main strength of the response 
lies at high energies. If we use the short-time expansion 
in Eq.\ (\ref{operator}) at this high $\Omega$ region,
 we can rewrite Eq. (\ref{timecorr}) for the probe 
$\qhat_q$ using the Baker-Campbell-Hausdorff formula as
\begin{equation}
    W(q,\Omega)\simeq {1\over 2\pi}\int\!\! dt\, e^{i(\Omega-q^2/2)t}
    \langle e^{-iq\hat{p}_xt+\cdots}\rangle_{gs},
\label{hausdorff}
\end{equation}
where the dots denote operators with higher powers of $t$. 
The expression shows that the response is peaked at the 
quasielastic energy $q^2/2$ and has a width increasing with $q$ and with, 
e.g., the momentum spread in the ground state. This holds precisely for 
a simple harmonic oscillator Hamiltonian, while in general is 
modified by anharmonicity effects.  The 
limiting form of response at large $q$ has been used to extract momentum 
distribution of complex system in terms of $y-$scaling 
analysis\cite{yscal}. In our case, as noted earlier, the model 
space of diagonalization limits the value of $q$ around 50 
with the corresponding limit $\sim 1/q$ in the resolution of the 
wave function. This is much smaller than the length 
parameter $1/\sqrt{\omega_0}$ of our oscillator basis. 
For the quartic oscillator the length scale will be 
modified from the simple oscillator value. One may define 
the characteristic length  scale  in the ground state by 
$\bar x_{gs}\equiv (\bra{gs}\hat x^2\ket{gs})^{1/2}$. 
Calculated values of $\bar x_{gs}^{-1}$ are 1.64 for $k=0.2$ and 
1.57 for $k=0.6$. Thus the operator at, say $q=20$, probes already a 
fairly fine structure of the system compared with the 
length scale of the ground state. 
The fact that the state ${\hat Q}_q|gs\rangle$ has an  oscillation
length scale $1/q$ also explains the occurrence of the quasielastic peak: 
The typical oscillation length 
scale of the harmonic oscillator wave function at energy $E\sim n\omega_0$ 
is $\sqrt{\langle{x^2}\rangle}/n\sim 1/\sqrt{E}$ which becomes 
$\sim 1/q$ in the region $E\sim q^2/2$.
Thus, the state ${\hat Q}_q|gs\rangle$ will have the largest overlap with
the states in the quasielastic region producing a peak in the response.

Let us now consider the sum rules. Low $n$ values of 
the sum $S_n\equiv S_n^{(0)}(\qhat_q)$ for the unsymmetrized probe 
are explicitly calculated 
to give
 $S_0=1,\,S_1=\half q^2, \, S_2={1\over 4}q^2 
(q^2 + {8\over 3}E_{gs})$, etc., where the last sum rule is 
obtained from the  virial theorem. For the symmetric probe the 
sum $\stil_n\equiv S_n^{(0)}(\qtil_q)$ 
is not analytically  obtained  but is given by the expectation 
values as: 
\begin{equation}
     \stil_0 = {1\over 4}\langle (1+\cos q(\xhat+\yhat)) 
                (1+\cos q(\xhat-\yhat))\rangle_{gs},\quad
     \stil_1 = {1\over 16}q^2\langle 2-\cos 2q\xhat
                 -\cos 2q\yhat)\rangle_{gs},\quad \hbox{etc.}. 
\label{stild01}
\end{equation}
It is useful to consider the limiting values for $q\rightarrow 0$ 
or $\infty$:  
\begin{eqnarray}
       & &\stil_0\rightarrow 1, \qquad 
               \stil_1\rightarrow {1\over 8}q^4
               \langle \xhat^2+\yhat^2\rangle_{gs} \quad : 
              \qquad {\rm for} \quad q\rightarrow 0, \\
       \label{stilgo0}        
       & &\stil_0\rightarrow {1\over 4}, \qquad 
             \stil_1\rightarrow {1\over 8}q^2 \qquad \qquad \qquad :
        \qquad {\rm for} \quad q\rightarrow \infty,
       \label{stilgoinf}
\end{eqnarray}
where the values at $q\rightarrow \infty$ are obtained under the 
assumption that the wavelength $1/q$ is much smaller than the 
typical length scale of the ground state wave function. These 
values are used to check the accuracy of the calculation within 
our model space, especially the one at large $q$ which 
requires the matrix elements Eq.(\ref{qelement}) with large $n$. 
Note that these values are almost independent of $k$, 
and the numerical calculation confirms 
that the $k$-dependence is small indeed. 
It turned out that the limiting values (\ref{stilgoinf}) for the sum 
rule are satisfied already at $q\simeq 10$. 
In view of Eq.\ (\ref{stild01}) this result implies that the ground state 
expectation values of $\cos q\xhat$, etc., are almost zero, 
i.e., the resolution at $q\simeq 10$ is already sufficiently 
fine for the ground state in accordance with the estimate given 
above.  For higher $n$ values 
the dependence of $\stil_n$ on $k$ is expected to become larger. 
Thus the gross structure of the response such as the total 
strength $\stil_0$ and average energy $\stil_1/\stil_0$ is 
rather insensitive to the values of $k$. Detailed structure 
related to high $n$ values of $\stil_n$ should reflect the dynamics. 

   Figure\ \ref{Respq}  shows the response 
function ${\tilde W}(q,\Omega)$ at 
$q=$ 10, 30 and 50. The gross structures 
at a given $q$ are similar for both $k=$0.2 and 0.6, and follow 
the behavior suggested earlier in this section: Not only the 
central energy follows $\Omega=q^2/2$ but the width of the response 
increases almost linearly with $q$. This should come out exactly 
from the sum rule if one had employed an unsymmetrized probe $\qhat_q$. 
Fine structure of the response, on the other hand, is quite 
different for the two cases. For  $k=0.2$ the response has a 
rather simple regular structure: at $q=$30, for instance, the response 
is  a superposition of a few structures with different sizes, 
each of which is centered around $q^2/2=450$ and is similar to 
the response of a harmonic oscillator given by
\begin{equation}
       W_{\hbox{ho}}(q,\Omega)=
       \sum_n\delta(\Omega-n\omega_0)f_n(\frac{q^2}{2\omega_0}),\qquad
       f_n(z)\equiv \frac{1}{n!}z^n e^{-z}.
\label{free}
\end{equation} 
In fact, by inspecting the wave functions 
one finds that these structures are related to the strong transition 
matrix elements of uncoupled (i.e., $k=0$) quartic oscillators 
which are integrable.  In contrast,  for $k=0.6$ 
the simple structure disappears and the values of the strength
change drastically from one state to the other, although one 
can still see even at $q=50$ a structure of the regularly spaced 
spikes as seen for $k=$0.2. 

For chaotic systems such as represented by random matrix theories, the 
amplitude distribution of wave functions or the matrix element 
distribution of an operator is known to follow the Porter-Thomas 
distribution. This may be contrasted to regular systems, where in 
many cases the quantum number imposes a selection rule of allowed 
transitions. Our result of response functions is in accordance with 
this generic behavior as far as the energy-strengths correlation 
is disregarded. In Fig.\ \ref{Dist} we show the strength distribution 
of the response function for $k=$0.2 and 0.6 at $q=$30. For $k=$0.6 the 
distribution follows the Porter-Thomas form  given by the dashed line and 
is quite different from that for $k=$0.2. 

The question then arises: What is the nature of the persisting regular 
structure in the response functions of Fig.\ref{Respq} for $k$=0.6. 
This structure becomes more visible if one introduces a 
normalized response function defined by 
\begin{equation}
   W_{\rm normalized}(q,\Omega)\equiv
   {W(q,\Omega)\over \tilde{W}(q,\Omega)}\bar{\rho}(\Omega),
\label{wnor}
\end{equation}
where $\bar{W}$ and $\bar{\rho}$ are respectively the response and the 
level density smoothed over energies. For the smoothing we 
employed the method of Strutinsky\cite{str} with the smoothing 
width of 20. This normalization 
procedure removes the gross structure effect of the 
response as constrained by the low order sum rules and enhances the 
embedded fine structure\cite{ams}. Figure\ \ref{resnor}  shows the 
normalized response for $k=$0.2 and 0.6. They 
show that the strengths in the regular spikes for $k=$0.2 are mostly 
redistributed for $k=$0.6 to produce smaller and smaller strengths to 
fill up the background, although one can still see the 
equidistant structure. The latter may be called an intermediate 
structure following Ref.\cite{lane}. This energy-strength correlation 
in the 
response function has been washed out in the strength distribution.  

The presence of an intermediate structure can be detected also in 
the response correlation function $C(\epsilon)$ defined by
\begin{equation}
C(\epsilon)\equiv\int dE~{\tilde W}(q,E)
                {\tilde W}(q,E+\epsilon).
\label{corr}
\end{equation}
For equidistant structures such as the one for the free response (\ref{free}), 
the correlation function gives again the regular  pattern with the same 
spacing, e.g.,
\begin{equation}
    C_{\hbox{free}}(\epsilon)=\sum_{n,n'}\delta(\epsilon-(n'-n)\omega_0)
    f_n(z)f_{n'}(z)=\int\!\frac{dt}{2\pi}\,e^{i\epsilon t}
    e^{2z(1-\cos\omega_0 t)},
\label{corfree}
\end{equation}
with $z\equiv q^2/2\omega_0$ and $f_n$ of Eq.(\ref{free}). 
Figure\ \ref{fig_corr} shows the correlation function
for the response function at $q=30$. In the actual calculation we used 
the normalized response function (\ref{wnor}) in order to remove the gross 
shape effect and used the level number displacement $\delta i$ instead of
the energy displacement $\epsilon$. 
The resultant correlation 
function was then smoothed with a smoothing width $\Delta i=4$.
We find the oscillator pattern arising from the intermediate
structure in the response function for both $k=0.2$ and 0.6.

To understand why there still remains the intermediate structure even 
in the chaotic case, let us investigate the nature of the 
peak levels in some detail. We first pick up the peak levels carrying 
the largest strength among neighboring levels in the response function. 
For $k=0.2$ the assignment of peak levels 
can be done without difficulty, while for 
$k=0.6$ there may be an ambiguity. The qualitative results of the
following analysis are however independent of this ambiguity. 

We first studied the NPC's of these peak levels in terms of the 
basis states (i.e., $\ket{j}=|n_1,n_2)_{\rm ES}$ in Eq.(\ref{basis})), 
and found that they are markedly smaller than NPC's for other 
levels. This implies that the mixing of the basis states in these 
peak levels is smaller than other states. The nature of the 
peak levels may become clear from Fig.\ \ref{nx-ny}, where we plot 
the quantity 
\begin{equation}
A_n^{(i)}\equiv \bra{i}|{\hat n}_x-{\hat n}_y|/({\hat n}_x
                 +{\hat n}_y)\ket{i}.
\label{defan}
\end {equation}
Here,  black points correspond to the peak levels 
and bars indicate the average values over neighboring levels. 
For $k=0.2$, the values of $A_n^{(i)}$ for peak levels
are nearly 0.9, almost twice of those for other levels. 
For $k=0.6$, although it is not so evident as for $k=0.2$,
the values for peak levels are larger than the average.
These facts strongly indicate that the peak levels are associated 
with the  basis states of the type $|0,m)_{\rm ES}$, although 
considerably affected by the mixing with other states for $k=$0.6. 
In fact, these basis states are the only states excited by
the probe ${\tilde Q}_q$ for a simple oscillator Hamiltonian. 

We may study the above results from the opposite direction. In 
Fig.\ref{lds} we show the NPC of Eq.(\ref{npc}) 
for the basis state $\ket{\alpha}\equiv|n_1,n_2)_{\rm ES}$ 
by taking the eigenstates $\ket{i}$ of the Hamiltonian  for 
the states $\ket{j}$ in Eq.(\ref{npc}). 
The abscissa is the basis number $\alpha$,  and the basis state type
$|0,m)_{\rm ES}$ is denoted by  crosses. 
For both $k$-values the basis states of the type $|0,m)_{\rm ES}$ 
have  the smallest NPC values, which shows that these basis states have
the smallest spreading width caused by the mixing with other basis states. 
Thus the fact that there remains an intermediate structure in the response 
function may be restated in the doorway state picture \cite{weisskopf}: 
The probe $\qtil_q$ excites first the doorway states $|0,m)_{\rm ES}$ 
which will then mix with other states causing 
the spreading of the  strengths.
As the spreading width for these specific states are smaller than the level 
spacing between the doorway states,  the intermediate structure emerges even 
in the chaotic case. 

Since the states $|0,m)_{\rm ES}$ may correspond to the classical isolated
periodic orbits along the $x$ and $y$ axis, the relation between the
intermediate structure in the response function and the scar \cite{heller} 
may be an interesting problem. 
Suppose that an initial wave packet $|\phi(0)\rangle$ is located at 
some point of the closed orbit having the period $T$. The  wave 
packet will then semiclassically evolve along the closed orbit and will 
return to the initial position at each time interval $T$. 
Accordingly, the overlap of the wave packet at time $t$ with the
initial wave packet $|\langle\phi(0)|\phi(t)\rangle|$ will have
peaks at $t=nT$, where $n$ is an integer. The value of these peaks 
will decay due to the instability of the closed orbit 
like $\exp(-\lambda/2t)$, where $\lambda$ denotes the Lyapunov exponent 
of the closed orbit\cite{heller}. Taking the state 
${\hat Q}_q|gs\rangle$ as the
initial wave packet $\ket{\phi(0)}$, the response 
function is nothing but the 
Fourier transform of the overlap $|\langle\phi(0)|\phi(t)\rangle|$. 
Therefore, the intermediate structure with peak level spacing $D=2\pi/T$
and the spreading width of the peaks $\gamma=\lambda$ may emerge 
in the response function if the condition $\gamma/D\le 1$ is satisfied. 
In the present model, the period of the 
closed orbits along the $x$- and $y$-
axes does not depend on the parameter $k$ and is given by 
$T={1 \over \sqrt{2\pi}} \Gamma({1 \over 4})^2\simeq 5.24$ at $E=1/2$. 
Numerical calculation also shows that the Lyapunov exponents for these 
closed orbits at $E=1/2$ are 
$\lambda\simeq 0.53$, and 0.92 for $k=0.2$ and $0.6$, respectively.
The values of $\gamma/D$ are then given by 0.44 and 0.77 for $k=0.2$ and
0.6, respectively, both of them being  smaller than unity. Note that
$\gamma/D$ is independent of the energy. Thus, the existence of the
intermediate structure may be explained also from the semiclassical
point of view.

\section{Summary}
\label{sec:summary}

In this paper we studied properties of the response functions for
a coupled quartic oscillator with several probes with a special
attention to the difference between the regular and the chaotic cases.

As a first example, we took the response to the probe ${\hat x^2}$.
Since the response function is determined by the
probe as well as the nature of the wave functions of the system,
we must pay attention also to the character of the probe.
For instance, the operators for which the diagonal matrix element
becomes a main component are not adequate to see the difference of the
dynamics.
For the operator ${\hat x^2}$, it can be decomposed  like Eq.\ (\ref{x2}),
and the operator $D_0$ has such a character,
while for the operators $D_2^\dagger$ and $D_2$ the non-diagonal
matrix elements are important.
Therefore, by removing the strengths associated with the operator
$D_0$ we can see the difference of the response between the chaotic and
the regular cases, namely more spreading of strengths for the chaotic case,
which can be quantified with the number of principal components.

Next, we considered the response to the probe 
${\hat Q}_q ={\rm e}^{{\rm i}q{\hat x}}$.
The response function at a given momentum transfer  $q$ is related to the 
time-correlation  function of the operator with a  resolution $1/q$ 
in the coordinate space.
It was shown that the gross structure of the response function
is similar for the chaotic and the regular cases as constrained by
global sum rules. On the other hand, the difference is reflected on
the fluctuation, as seen in the strength distribution ( the histogram
of strengths). Moreover, we detected the intermediate structure
(i.e., typical energy scales) even in the chaotic case which can not
be expected for the random-matrix model. We found that the existence of
the intermediate structure is due to the fact that the spreading
width of the doorway states is smaller than the level spacing of the 
doorway states, and also indicated its relation to the scar phenomenon.
It would be interesting to study this structure from a 
different point of view, e.g., the semiclassical theory of 
responses \cite{Wilkinson,eckardt,MW,mehlig} based on the periodic orbits.

In this paper for the sake of simplicity we restricted the 
discussion to the transitions 
between the states belonging to the same symmetry class.
In the realistic situation, however, the transition connecting states with 
different symmetry classes may also occur at same time.
It is well known that the level spacing statistics 
drastically changes when we consider the levels belonging to
different symmetry classes simultaneously. 
Thus, it is also interesting to see what happens for the
response function to the probe connecting different symmetry classes.

\vspace{6mm}
The authors thank M.Matuso for valuable discussions. They thank also 
P.Schuck for a discussion about the semiclassical description of 
response functions. 

 
\begin{figure}
 \caption{ Response to the probe ${\tilde {x^2}}$ for the initial 
     state $\ket{i=500}$ as a function of the  energy 
     $E$ for $k=$0.2 (a) and 0.6 (b). Contained levels in the displayed
     energy regions are 459th to 542th for $k=0.2$ and
     462th to 554th for $k=0.6$.}
\label{Respx}
\end{figure} 

\begin{figure}
\caption{Response to the operator $D_0$
     for the initial 
     state $\ket{i=500}$ as a function of   energy 
     $E$ for $k=$0.2 (a) and 0.6 (b), and the one for the operator
     $D_2^\dagger$ for $k=0.2$ (c) and 0.6 (d). 
}
\label{resd0d2}
\end{figure}

\begin{figure}
\caption{The NPC $N_{\rm pc}^{(i)}$ for the state
$D_0\ket{i}$ as a function of the energy
of the initial eigenstate $E_i$ for $k=0.2$ (a) and 0.6 (b).
}
\label{d0npc}
\end{figure}
 
\begin{figure}
\caption{Same as Fig.\protect\ \protect\ref{d0npc} but for $D_2^\dagger$.
}
\label{d2npc}
\end{figure}

\begin{figure}
  \caption{ The fraction of the ${\tilde {x^2}}$ strengths 
     (omitting the one for $E=E_1$) carried by 
     the two major states to the total strengths is plotted 
     for each initial state $\ket{i}$ as a
     function of the initial state energy $E_i$.  }
\label{Strength}
\end{figure}

\begin{figure}
  \caption{ Response function ${\tilde W}(q,E)$ for  $k=0.2$ 
  at $q=$ 10 (a), 30 (c), and 50 (e)
          and for $k=0.6$ at $q=$ 10 (b), 30 (d), and 50 (f).
        Note the changes in the scale of the vertical axis.}
\label{Respq}
\end{figure} 

\begin{figure}
  \caption{Strength distribution $P(S^{1/2})$ of the matrix element 
  $S\equiv |\bra{j}{\tilde Q}_q\ket{gs}|^2$ 
  at $q=30$ for $k=0.2$ (a) and 0.6 (b). Strengths are normalized
  as Eq.\protect\ (\protect\ref{wnor}). Dashed line shows the Porter-
  Thomas distribution.
  }
\label{Dist}
\end{figure}

\begin{figure}
  \caption{The normalized 
  response function Eq.\protect\ (\protect\ref{wnor}) at $q=$30
  for  $k=$0.2 (a) and  0.6 (b).
  Compare with the responses at $q=$30 shown in 
  Fig.\protect\ \protect\ref{Respq}.
  }
\label{resnor}
\end{figure}

\begin{figure}
\caption{Smoothed correlation function $C(\delta i)$ for the normalized 
response function to the probe $\qtil_q$ with $q=30$ as a function
of the level number displacement $\delta i$ for $k=0.2$ (a) and
0.6 (b). Smoothing width $\Delta i=4$ is adopted.
}
\label{fig_corr}
\end{figure}

\begin{figure}
\caption{The quantity $A_n^{(i)}$ for $k=0.2$ (a) and 0.6 (b).
Black points denote the peak levels for the probe ${\tilde Q}_q$ with $q=30$ 
 and bars show the average
over neighboring levels.
}
\label{nx-ny}
\end{figure}

\begin{figure}
\caption{The NPC $N_{\rm pc}^{(\alpha)}$ for basis state $\ket{\alpha}$
belonging to the symmetry class $A_1$ for $k=0.2$ (a) and 0.6 (b).
Horizontal axis shows the level number of the basis state $\ket{\alpha}$.
Cross points correspond to the basis states type $|0,m)_{\rm ES}$.
}
\label{lds}
\end{figure}

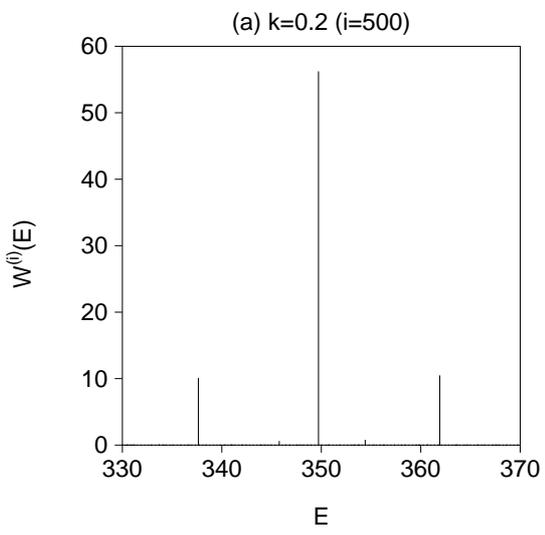
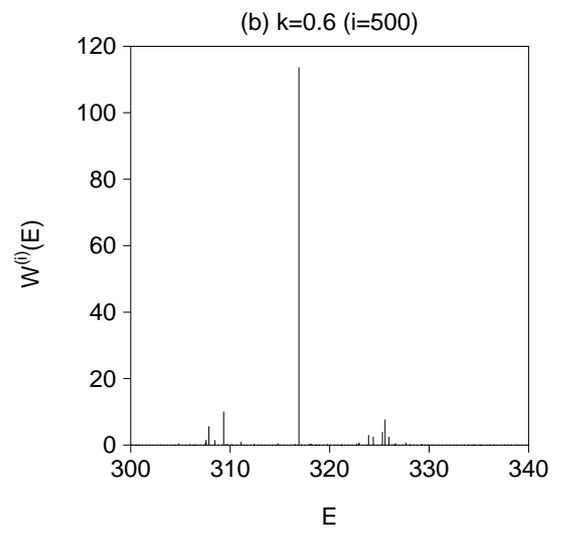
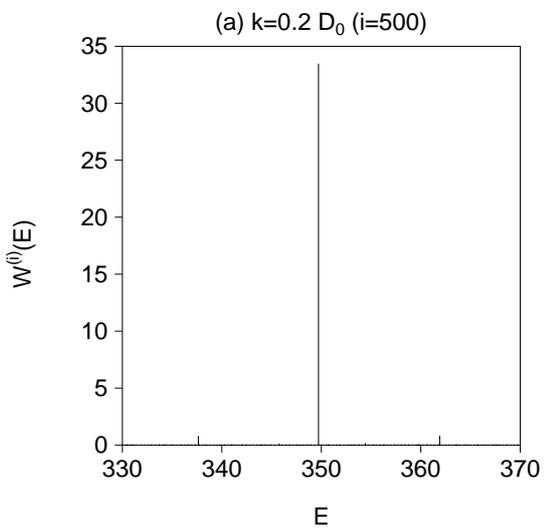
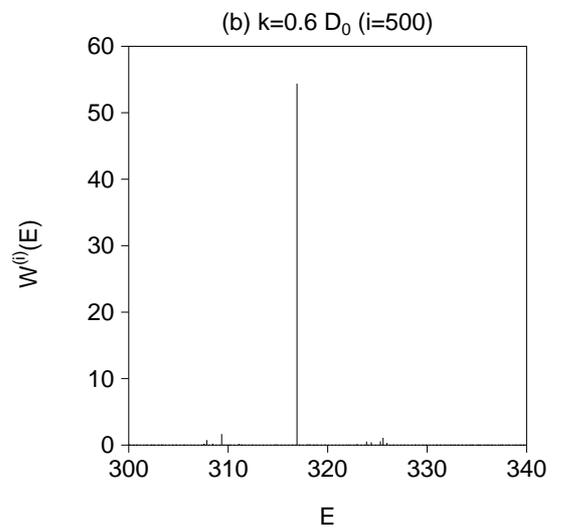
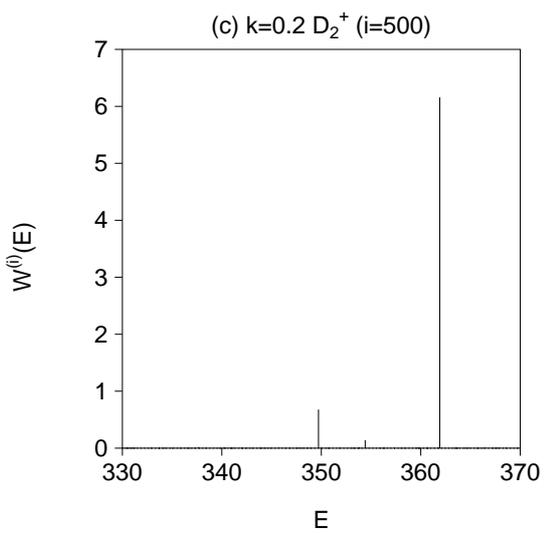
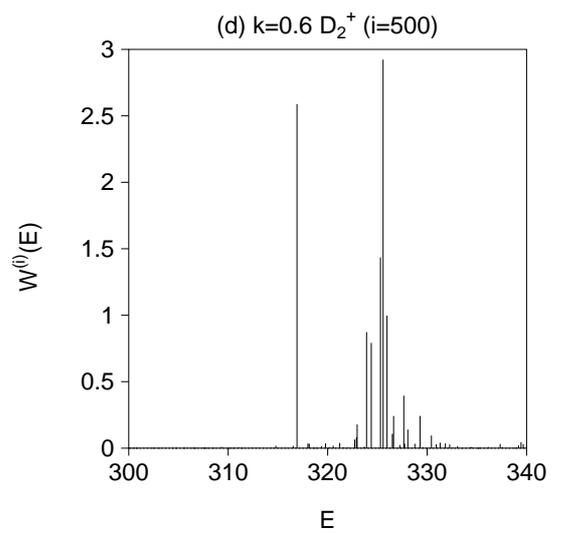
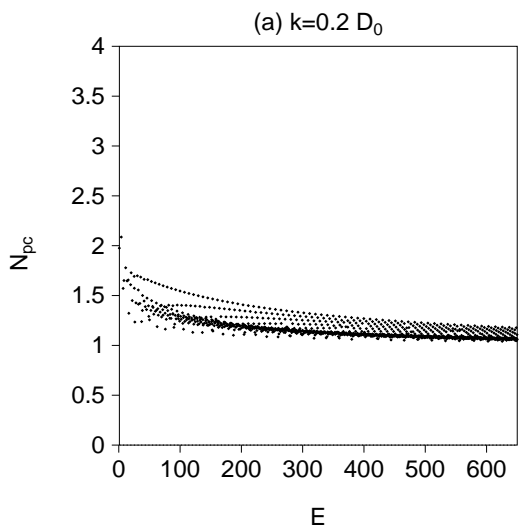
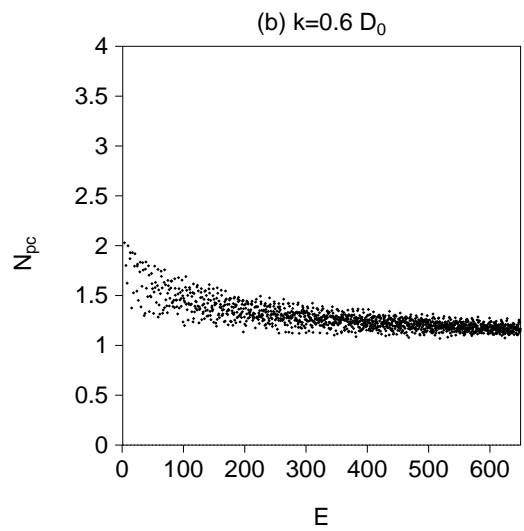
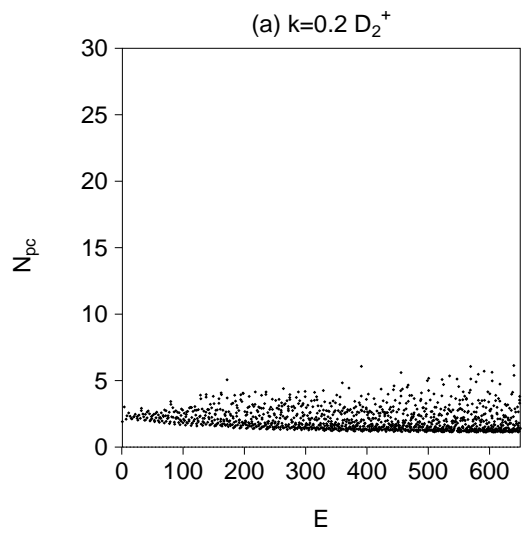
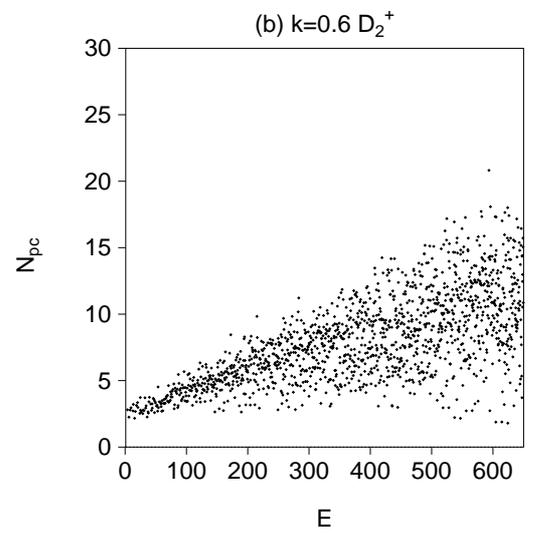
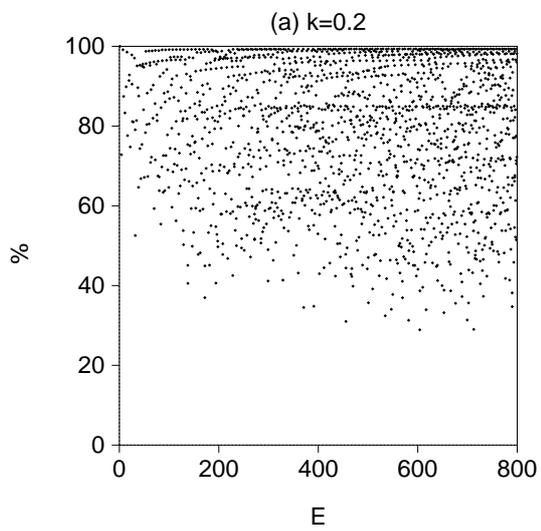
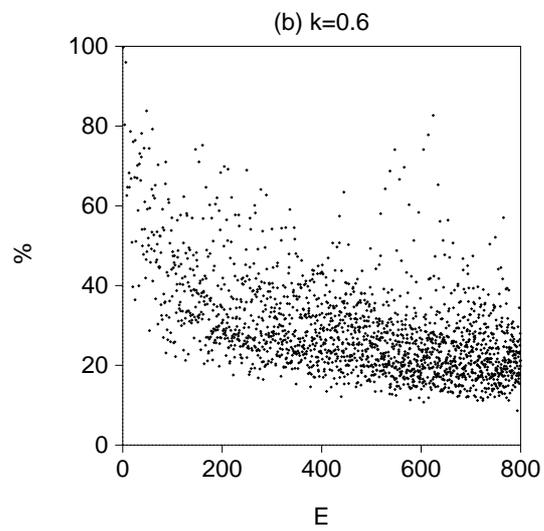
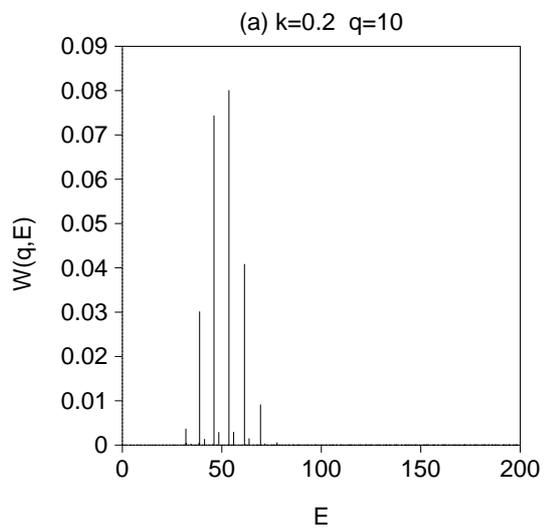
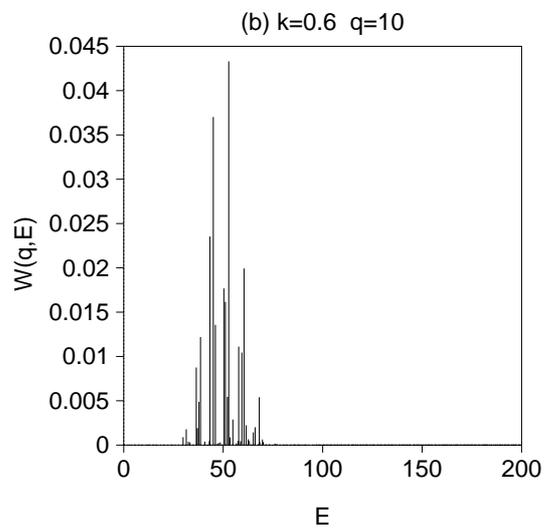
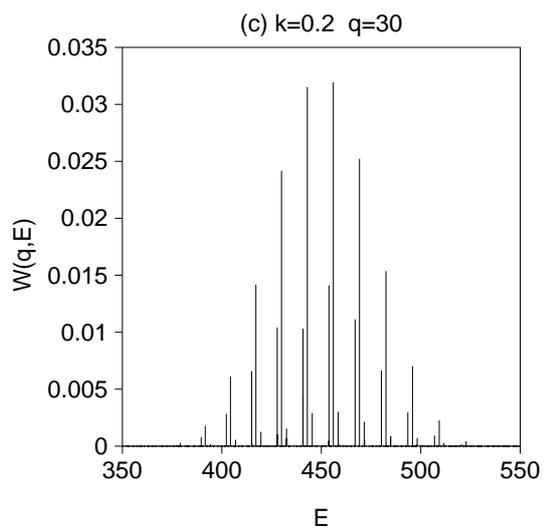
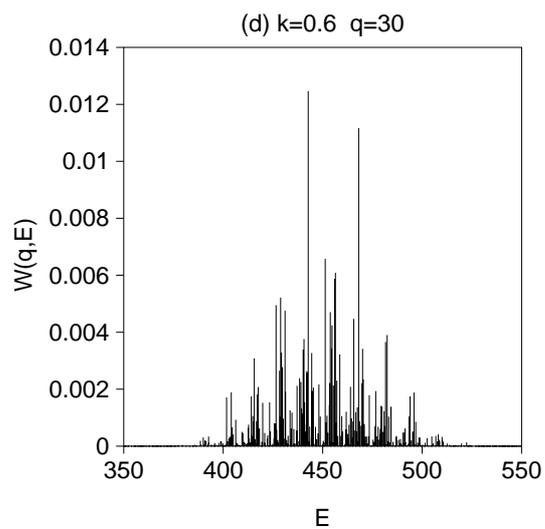
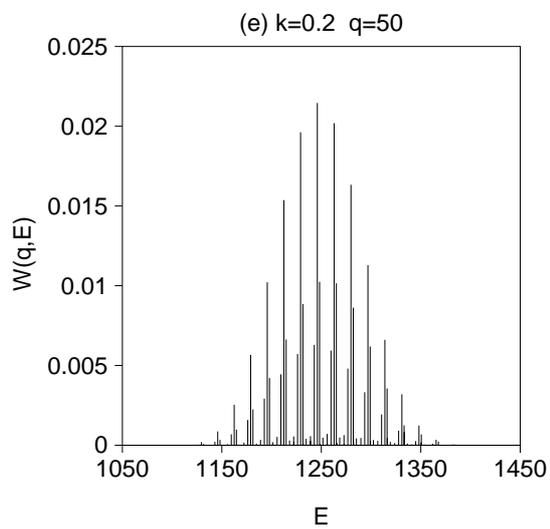
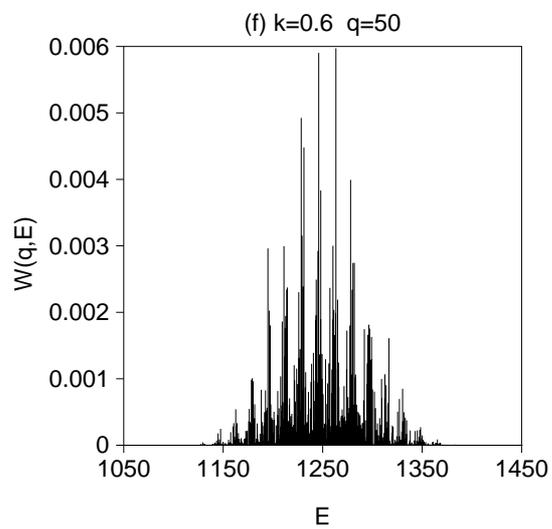
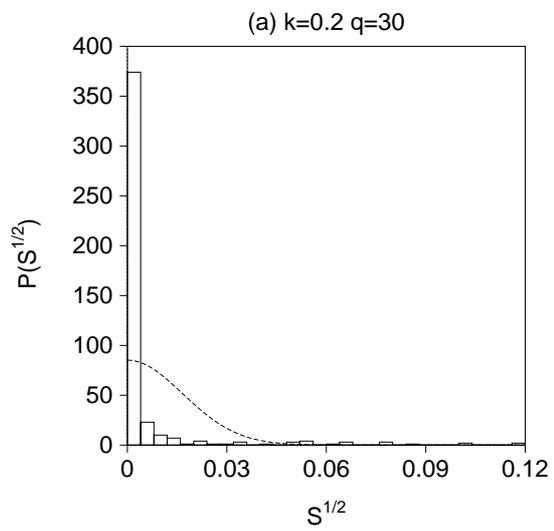
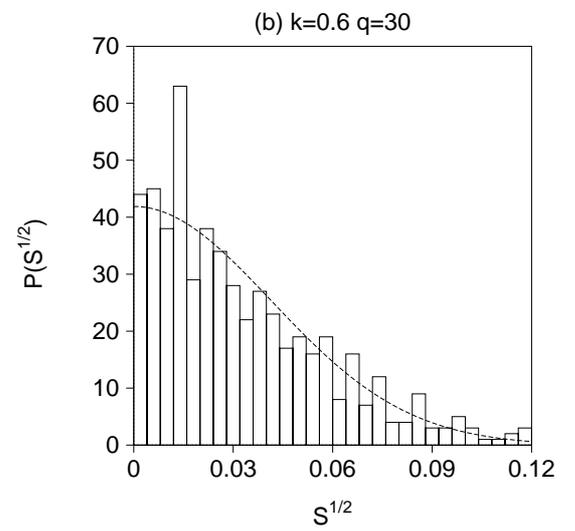
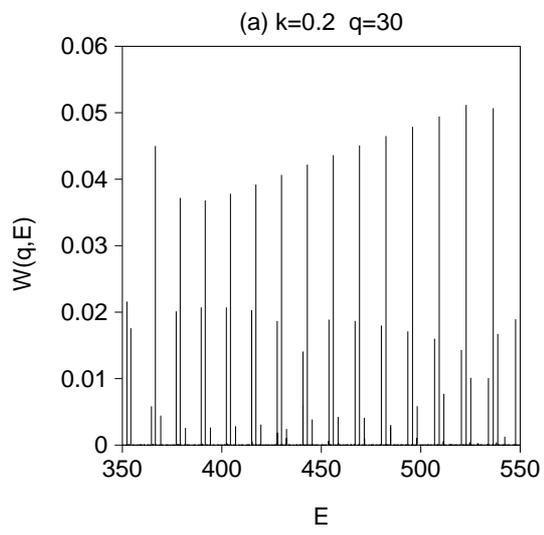
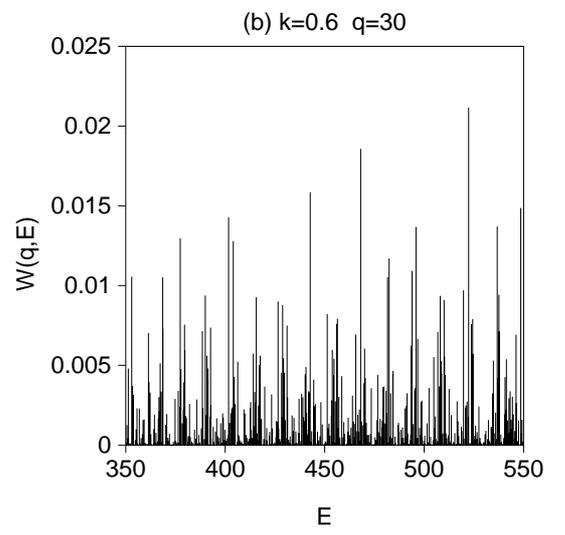
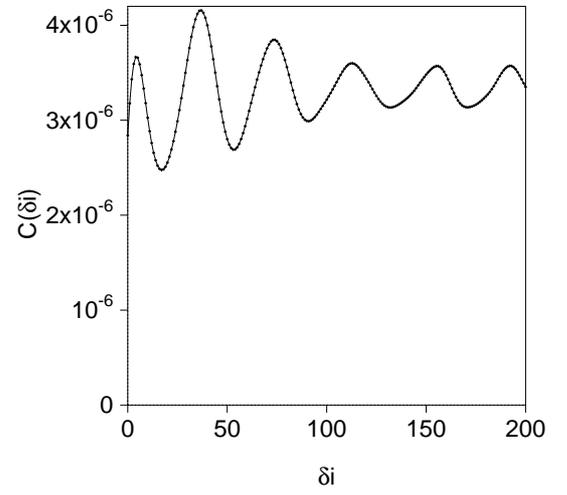
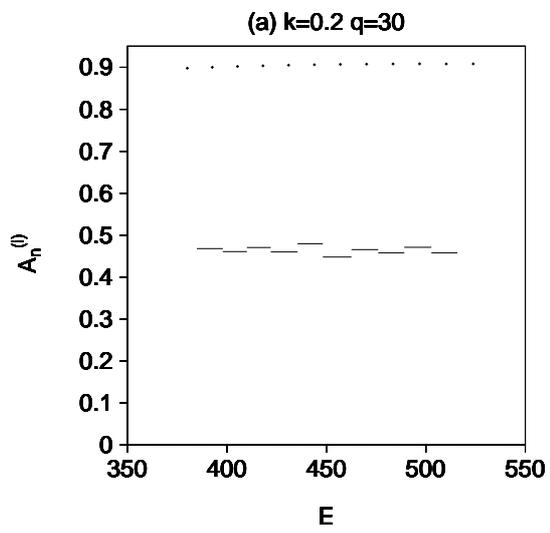
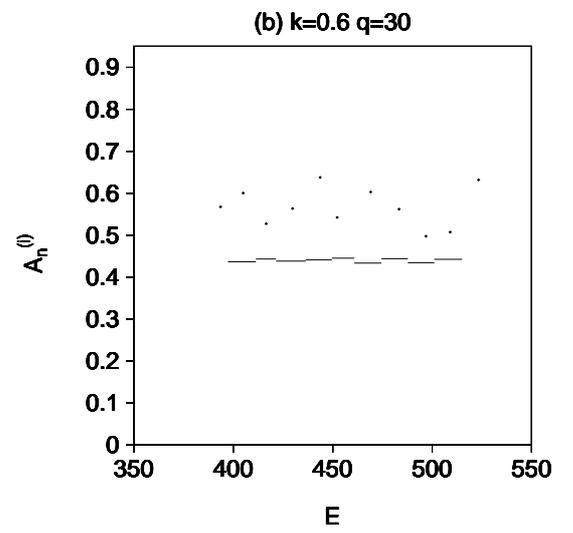
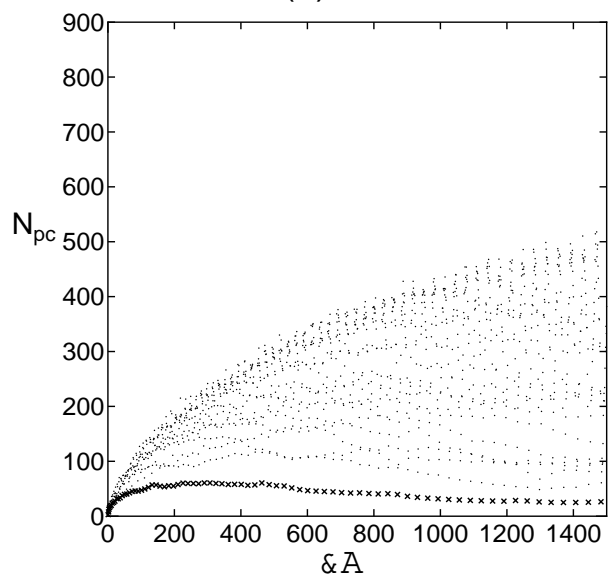
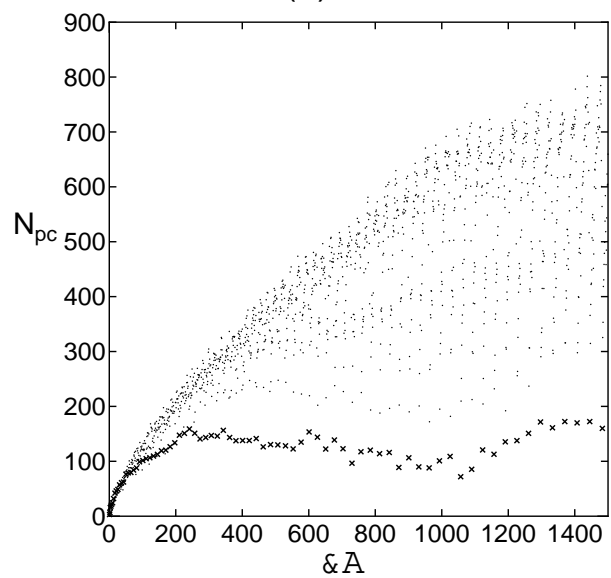
\end{document}